\documentclass[reprint,twocolumn,preprintnumbers,amsmath,amssymb,groupedaddress,superscriptaddress,prb]{revtex4-1}
\usepackage{amsmath,amsfonts,amssymb,graphicx,color,times,bbm}
\usepackage{graphicx}
\usepackage{dcolumn}
\usepackage{bm}
\usepackage{amssymb}
\usepackage{amsmath}
\usepackage[colorlinks=true,citecolor=blue,urlcolor=blue]{hyperref}

\usepackage{footmisc}
\usepackage{braket}
\usepackage{multirow}
\renewcommand{\emph}{\textit}
\graphicspath{{./}}
\begin{document}

\title{Noise spectroscopy of a quantum-classical environment with a diamond qubit}

\author{S. Hern\'{a}ndez-G\'{o}mez}
\affiliation{LENS European Laboratory for Non linear Spectroscopy, Universit\`a di Firenze, I-50019 Sesto Fiorentino, Italy} 
\affiliation{CNR-INO Istituto Nazionale di Ottica del Consiglio Nazionale delle Ricerche, I-50019 Sesto Fiorentino, Italy}
\author{F. Poggiali}
\affiliation{LENS European Laboratory for Non linear Spectroscopy, Universit\`a di Firenze, I-50019 Sesto Fiorentino, Italy} 
\affiliation{CNR-INO Istituto Nazionale di Ottica del Consiglio Nazionale delle Ricerche, I-50019 Sesto Fiorentino, Italy}
\author{P. Cappellaro}
\affiliation{LENS European Laboratory for Non linear Spectroscopy, Universit\`a di Firenze, I-50019 Sesto Fiorentino, Italy} 
\affiliation{Department of Nuclear Science and Engineering, Massachusetts Institute of Technology, Cambridge, MA 02139}  
\author{N. Fabbri}\email{fabbri@lens.unifi.it}
\affiliation{LENS European Laboratory for Non linear Spectroscopy, Universit\`a di Firenze, I-50019 Sesto Fiorentino, Italy} 
\affiliation{CNR-INO Istituto Nazionale di Ottica del Consiglio Nazionale delle Ricerche, I-50019 Sesto Fiorentino, Italy}

\begin{abstract}
Knowing a quantum system's environment is critical for its practical use as a quantum device. Qubit sensors can reconstruct the noise spectral density of a classical bath, provided long enough coherence time. Here we present a protocol that can unravel the characteristics of a more complex environment, comprising both unknown coherently coupled quantum systems, and a larger quantum bath that can be modeled as a classical stochastic field. We exploit the rich environment of a Nitrogen-Vacancy center in diamond, tuning the environment behavior with a bias magnetic field, to experimentally demonstrate our method.  We show how to reconstruct the noise spectral density even when limited by relatively short coherence times, and identify the local spin environment. Importantly, we demonstrate that the reconstructed model can have predictive power, describing the spin qubit dynamics under  control sequences not used for noise spectroscopy, a feature critical for building robust quantum devices. At lower bias fields, where the effects of the quantum nature of the bath are more pronounced, we find that more than a single classical noise model are needed to properly  describe the spin coherence under different controls, due to the back action of the qubit onto the bath.
\end{abstract}
\maketitle

\section{ Introduction }
Characterizing the interaction of a qubit with its environment is critical to realize robust  quantum devices. A full understanding of the qubit environment enables developing effective strategies against decoherence, including optimized dynamical decoupling (DD) sequences~\cite{Poggiali18,Uhrig07} and quantum error correction codes~\cite{Layden18}.  Moreover, part of the environment might display coherent coupling to the  qubit and thus provide an additional resource to enhance  its computational or sensing performance~\cite{Goldstein11,Abobeih18,Cooper18x}. 

Fortunately, the qubit itself is  a sensitive probe of its local environment. 
In addition to  $T_2^*$ relaxometry~\cite{Myers14,Rosskopf14,Vandersar15,Stark17} 
and spin locking schemes~\cite{Bylander11,Belthangady13}, the most common and powerful noise spectroscopy methods~\cite{Almog11,Yuge11,Alvarez11,Kotler11,Young12} rely on the systematic analysis of the sensor decoherence under sets of DD control sequences~\cite{Viola98,Cywinski08,Faoro04,Biercuk11}.  
Periodic DD sequences realize narrow frequency filters that select only a specific noise contribution, while canceling all  other interactions. This method has been used for noise identification  with spin qubits in diamond~\cite{Reinhard12,Bar-Gill12,Romach15}, superconductive flux qubits~\cite{Yoshihara14}, trapped ions~\cite{Kotler13}, and nanoelectronic devices~\cite{Muhonen14}. While the filter function approach has been extended in some cases to more complex and quantum baths~\cite{Uhrig07,Paz-Silva13}, 
most of these noise spectroscopy methods usually assume the environment to be a classical stochastic bath~\cite{Alvarez11a,Biercuk11,Kotler11}. In addition these methods rely on the assumption that the noise is weak enough to allow relatively long qubit coherence time under the applied control. 

\begin{figure}[t!]
\includegraphics[width=\columnwidth]{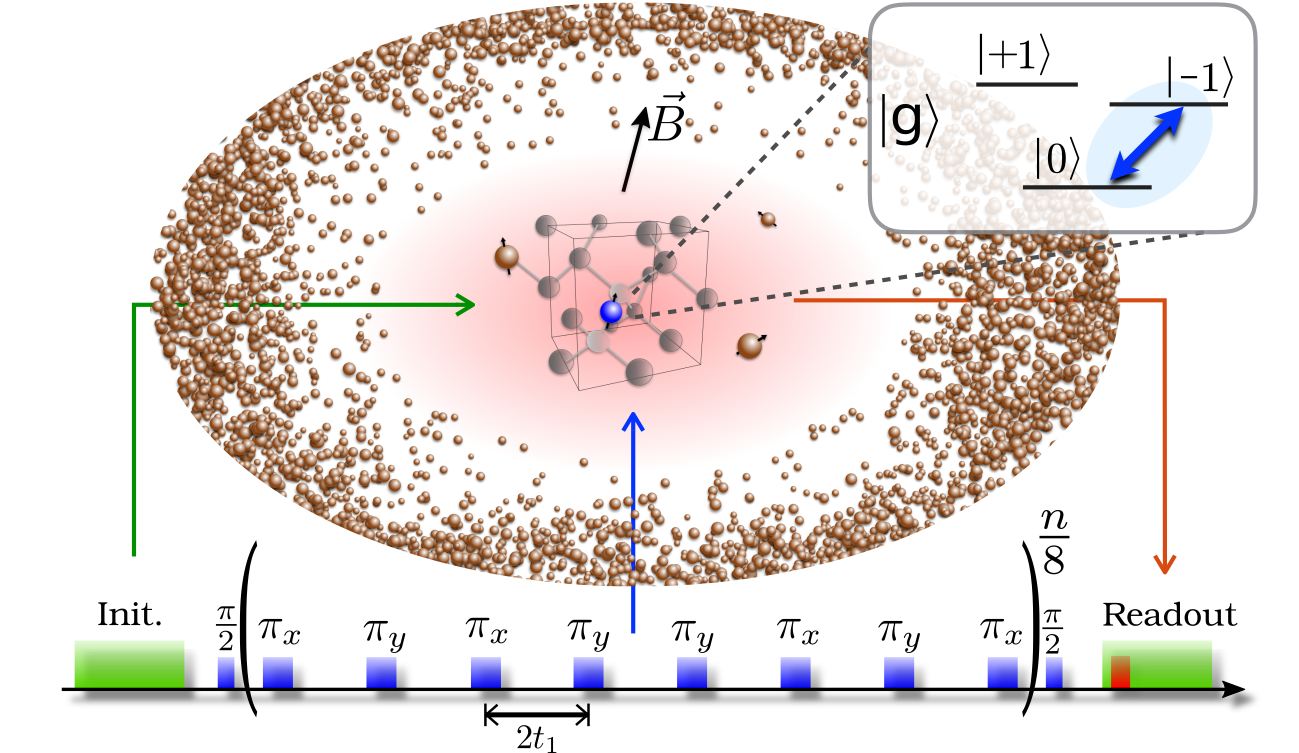}
\caption{\textbf{System-environment model and experimental protocol.} The NV electronic spin (blue sphere) is sensitive to $^{13}$C impurities in the diamond (brown spheres), including isolated nearby spins and a larger ensemble spin bath. The NV is addressed and manipulated via optically-detected magnetic resonance in the presence of an external bias magnetic field,  aligned with the NV axis. After optical initialization  in the $m_s = 0$ spin  state, the NV is manipulated with  resonant microwave pulses (in blue), and read out optically (red). A detailed description of our experimental setup can be found in~\cite{Poggiali18,Poggiali17, Suppl}.
}
\label{fig:exp}
\end{figure}

\begin{figure*}
\includegraphics[width=\textwidth]{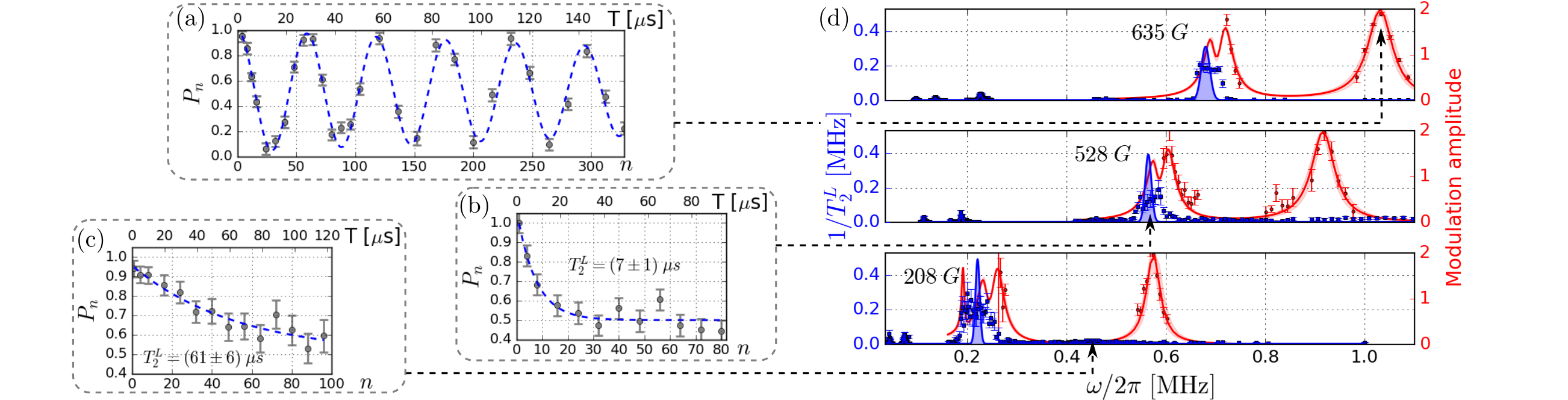}
\caption{(a-c) Spin coherence, mapped onto population $P_n$, as a function of $n$.  The inter-pulse delay time $t_1$ is fixed: (a)  $t_1=242$~ns, $B=635$~G;  (b) $t_1=456$~ns, $B=528$~G; (c) $t_1=585$~ns, $B=208$~G. (d) NSD and coupling with nearby nuclei. We report $1/T_2^L$ (blue, left-hand side vertical scale) and the amplitude of the observed coherent modulations (red, right-hand side vertical scale), for different magnetic field strengths ($B=208-635$~G). Blue lines are the fit of high-order harmonics to extract the NSD, and red lines are the simulations of the completely resolved nearby Carbons (see text). The red shadow describes the uncertainty on the estimation of the coupling strength components. The coupling strength to the third carbon is too weak to be distinguishable from the $0^{th}$-order collapse, thus it has been extracted from higher-order harmonics (see Fig.~III.S~\cite{Suppl}).}
\label{fig:coherenceN}
\end{figure*}

Here, we experimentally demonstrate a protocol for characterizing the qubit environment that overcomes the challenges arising when those assumptions are not verified. We implement the protocol using the electron spin qubit associated with a single Nitrogen-vacancy (NV) center in diamond, which  has emerged as a powerful platform for quantum technologies~\cite{Doherty13,Rondin14}. The NV qubit displays a complex environment, comprising $^{13}$C nuclear spins randomly distributed in the diamond lattice. The thermal and quantum fluctuations of this environment, and the distribution of environment-qubit interaction strengths  make this an extremely rich scenario where to test our protocol. The environment can be divided into a small set of resolved $^{13}$C, and a large ensemble of unresolved $^{13}$C that we treat as a collective bath  (Fig.~\ref{fig:exp}). We can further tune the ratio between the environment internal energy and its coupling to the NV center by varying the strength of an applied external magnetic field, thus exploring different bath regimes~\cite{Reinhard12}. 

Crucially for quantum devices, we show that the acquired knowledge of a classical (weakly-coupled) bath can reliably predict the qubit dynamics  even under drivings  that differ from the ones used for noise spectroscopy. 
Conversely, we find that the assumption of one simple classical model describing an intrinsically quantum bath, strongly coupled to the qubit, is not always appropriate to achieve a predictive model for all dynamics. Instead, distinct classical models of the spin bath are needed to predict the qubit spin behavior under different control schemes, reflecting that the bath feels the qubit back action, which varies with the control sequences driving the qubit dynamics.

\section{{Experimental system} }
We investigate the environment of a single deep NV center in an electronic-grade bulk diamond (Element6), with nitrogen concentration $[^{14}$N$] < 5$ ppb, and natural abundance of $^{13}$C (nuclear spin $I=1/2$). 
In the presence of a static bias field $B$ aligned along the NV axis, we can restrict the description to an NV spin subspace, $\{\ket{0}$, $\ket{-1}\}$. The system-environment Hamiltonian is 
\begin{equation}
\mathcal{H}=\omega_0 \sigma_z^{{\mbox{\tiny{(NV)}}}}+\frac{\omega_L}{2}\sum_k\sigma_z^{{\mbox{\tiny{($k$)}}}}+ \hbar\sum_k\sigma_z^{{\mbox{\tiny{(NV)}}}} {\bm \omega}_h^{{\mbox{\tiny{($k$)}}}}\cdot {\bm \sigma}^{{\mbox{\tiny{($k$)}}}},
\label{eq:ham}
\end{equation}
where $\omega_0=\gamma_e B$ and $\omega_L=\gamma_n B$, with $\gamma_e$ and $\gamma_n$ being the electron and nuclear gyromagnetic ratios, and $\omega_h^{{\mbox{\tiny{($n$)}}}}$ the hyperfine-interaction frequency tensor. The last term incorporates a small set of discrete couplings that can be fully resolved, as later shown, and a broad unresolved distribution of couplings that we describe as a collective bath. In the strong coupling regime, where the typical coupling strength overcomes the environment internal energy ($\|\omega_h\|\geq\omega_L$), the creation of entanglement between spin qubit and a large environment, with subsequent tracing over of the environment, induces loss of qubit coherence. In the weak coupling limit, $\|\omega_h\|\ll\omega_L$, the environment can be modeled as a classical stochastic field, as described in the Supplemental Material~\cite{Suppl},  also leading to a non-unitary qubit dynamics (dephasing). 

To characterize the spin environment, we reconstruct the environment-induced NV dynamics under sets of resonant multipulse control~\cite{Yuge11,Alvarez11}. 
The control field acting on the spin qubit can be described by a modulation function $y_n(t)$ and its squared Fourier transform  defines the  filter function $Y_n(\omega)$~\cite{Biercuk11}. Due to the presence of the bath, coherence decays as $W(t)= e^{-\chi(t)}$, where $\chi(t)$ depends on the noise spectral density (NSD) $S(\omega)$, as 
\begin{equation}
\chi(t)=\int\frac{d\omega}{\pi\omega^2}S(\omega)|Y(\omega)|^2.
\label{eq:chi}
\end{equation}
To measure $S(\omega)$, we perform a systematic spectral analysis of coherence under DD sequences of equispaced $\pi$ pulses, with increasing number of pulses. We use the XY-$8$ sequence~\cite{Gullion90} as a base cycle (Fig.~\ref{fig:exp}), as it is designed to improve robustness against detuning and imperfections of the $\pi$-pulse shape. The DD sequences are incorporated in a Ramsey interferometer that maps residual coherence after $n$ pulses into the observable population of the $\ket{-1}$ state, $P_n=(1+W)/2$.

\section{Environment spectroscopy} 
\label{Sec:spectroscopy}
\textit{Collective bath.} For long enough evolution time ({\it i.e.}, large number of pulses $n$), equispaced sequences with  interpulse delay $2t_1$ are well described by narrow monochromatic filters given by $\delta$-functions centered at $\omega=\pi/2t_1$. 
In this limit, $\chi$ depends only on the NSD spectral weight  at that specific frequency, whereas  all  nearby noise components are filtered out. Then, varying the number of pulses at fixed $t_1$
the coherence is expected to show an exponential decay, with a generalized coherence time $T_2^L$~\cite{Yuge11}
 \begin{equation}
W(nt_1) = \exp\left(-\frac{2 n t_1}{T_2^L}\right)\quad\textrm{with}\quad  S(\pi/2 t_1)\simeq\frac{\pi^2}{8 \,T_2^L }
\label{eq:expt2L}.
\end{equation}
The decay is faster for $t_1$ corresponding to the spin bath characteristic frequencies (coherence collapses), as shown in Fig.~\ref{fig:coherenceN} (b-c). 
Then, a practical protocol would be to map out $T_2^L$ varying $t_1$ around the first collapse, a region that carries the most information about the NSD. 
Fig.~\ref{fig:coherenceN}(d) shows $1/T_2^L$ as a function of $\omega=\pi/2t_1$ (blue dots).
\begin{figure}[thb]
\includegraphics[width=\columnwidth]{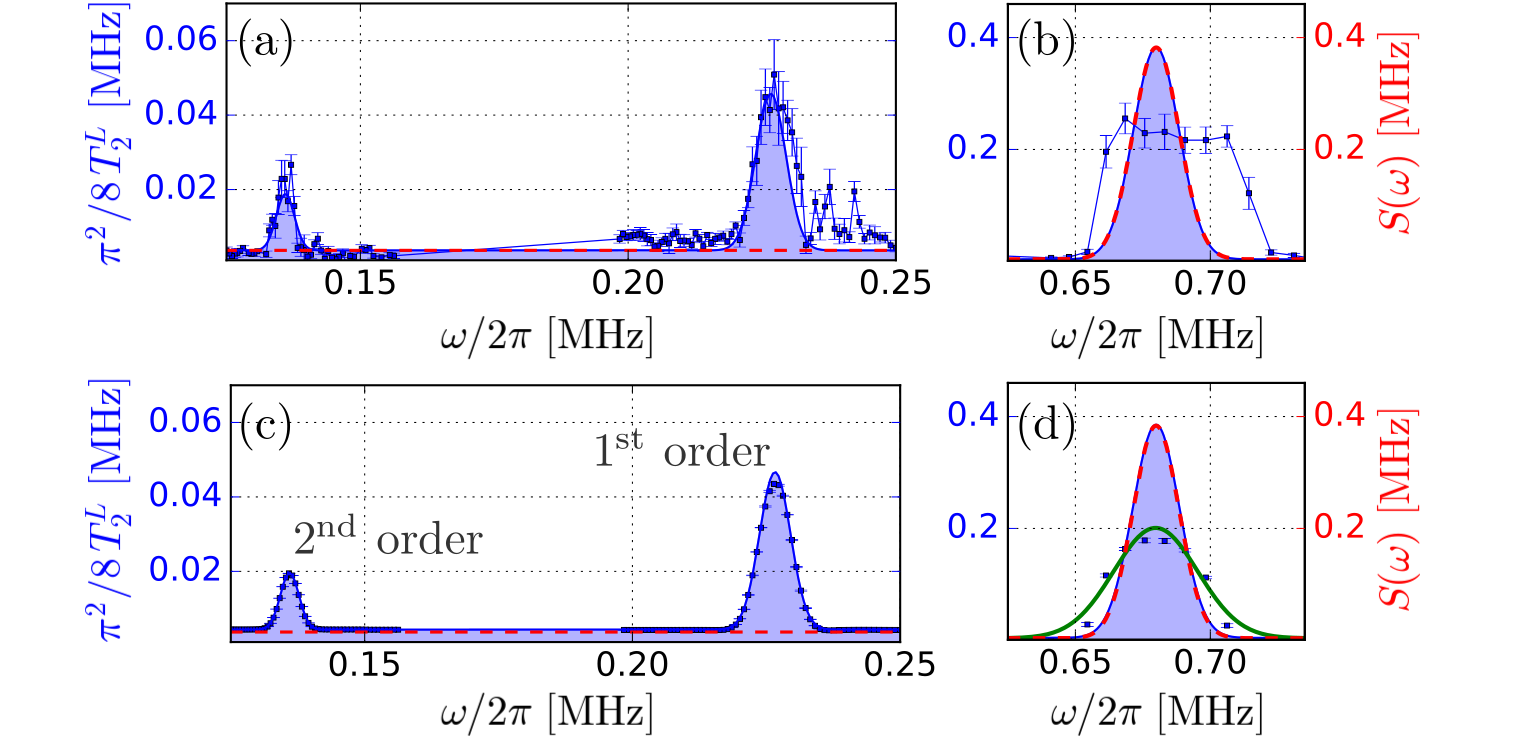}
\caption{$1/T_2^L$ and $S(\omega)$. (a-b)  Dots are experimental $1/T_2^L$ values peaked (a) at $\omega_L/5$ ($l=1$) and $\omega_L/3$ ($l=2$), and (b) at  $\omega_L$ ($l=0$), with $B=635(1)$~G. The blue line is a Gaussian fit of  harmonics  $l=1,2$, from which we extract $S(\omega)$ (red dashed line, peaked at $\omega_L$). (c-d) Reconstruction of a model NSD, used to proof self-consistency of the  method. The blue squares are $1/T_2^L$ values resulting from the fit of the simulated coherence. The solid lines are the Gaussian fits to the harmonics of orders $l=1,2$ (blue) and $l=0$ (green). The red dashed line is the original model NSD.}
\label{fig:harmonics}
\end{figure}

However, a strongly-coupled spin bath may lead to very fast decay (already at $n<8$) for $t_1$ around the first collapse, so that using large number of pulses is not possible. Unfortunately, for low number of pulses the filter induces an additional broadening of the NSD.
Since a sequence of equidistant $\pi$ pulses acts on the spin evolution as a step modulation function $y_n(t)$ with periodic sign switches, the filter function $Y_n(\omega)$ is not a single $\delta$-function, but shows periodic sinc-shaped peaks at frequency $\omega_l=(2l+1)\omega$, which can be well approximated by a periodic comb of $\delta$-functions only for large $n$. 
Then, $T_2^L(\omega)$ is affected not only by $S(\omega)$, but also by its higher harmonics~\cite{Yuge11,Alvarez11},
\begin{equation}
\frac{1}{T_2^L(\omega)}= \frac{8}{\pi^2} \sum_{l=0}^{\infty} \frac{1}{(2l+1)^2} S(\omega_l),
\label{eq:NSD_Fomega}
\end{equation}
giving the approximation in Eq.~\ref{eq:expt2L} for $l=0$. 

This last observation gives us a simple tool to  overcome the limitation of the short coherence decay time  in the collapses time windows: We center the higher order harmonics of the filter function around the expected NSD peak and combine the information  from several harmonics. This partially attenuates strong noise that would saturate the coherence decay and achieves a better approximation to a $\delta$-function, as for fixed number of pulses, the filter function gets narrower at higher orders. 

To validate our protocol, we  simulate $W(nt_1)$ under a simple noise model (a Gaussian centered at $\omega_L\!=\!2\pi\!\times\!750$~kHz), and  verify that $1/T_2^L$ obtained from the $0^{th}$-order filter harmonics ($l=0$) exhibits significant disagreement with the original spectrum, whereas 
the $l=1,2$ harmonics are sufficient to fully reconstruct the NSD peak (Fig.~\ref{fig:harmonics}(c)-(d), and Table~I.S~\cite{Suppl}). In experiments, we extract the NSD lineshape from a Gaussian fit of $1/T_2^L$ around first and second order collapses (Fig.~\ref{fig:harmonics}(a)).
Figure~\ref{fig:harmonics}(b) shows the obtained NSD lineshape (red dashed line), compared with data from the $0^{th}$-order collapse~\footnote{We attribute the extra broadening of $0^{th}$-order experimental data, compared with simulation, to the presence of the least-strongly-coupled among the observed nearby nuclei. This Carbon is clearly visible in higher-order harmonics, while is not resolved in the $0^{th}$-order collapse, where induces an overestimation of the NSD width.}.

\textit{Resolved nuclear spins.} The reconstruction of the NSD from a classical model fails in some narrow time windows, where the coherence  presents sharp dips  reaching even negative values ($P_n<0.5$). This  allows us to identify the coherent coupling of the NV spin  to a local small quantum environment, which becomes visible as the equispaced DD sequences partially filter out the larger spin bath.
The hyperfine interaction to single proximal nuclear spins (Eq.~\ref{eq:ham})  induces different phases  for the two states $|\pm\rangle=(|0\rangle\pm|1\rangle)/\sqrt{2}$ during the spin evolution time~\cite{Taminiau12,Kolkowitz12l,Zhao12}. The residual NV spin coherence then presents coherent modulations as a function of the pulse number [Fig.~\ref{fig:coherenceN}(a)], which 
give information on the hyperfine coupling tensor between the NV spin and nearby nuclear spins. 
The modulation amplitude shows sharp peaks as a function of frequency [red dots in  Fig.~\ref{fig:coherenceN}(d)], from which we can identify three different $^{13}$C nuclei and obtain an estimate of the energy-conserving component of the coupling strength, $\omega_h^{||}$~\cite{Taminiau12}. By fitting the  modulations with a periodic function $M_n(T)$~\cite{Suppl} as shown in Fig.~\ref{fig:coherenceN}(a), we extract a refined estimate of the parallel and orthogonal components of the coupling strength (see Table~III.S~\cite{Suppl}).
Note that we treat each coherently coupled $^{13}$C spin separately, since intra-spin couplings are negligible. A more detailed analysis (see e.g. refs.~\cite{Zhao11,Abobeih18}) could be used to identify as well  couplings between nuclear spins if they are strong enough to affect the dynamics. 

Characterizing the spin environment of a qubit is critical to achieve improved error correction protocols. One could, e.g., exploit the coherently coupled nuclear spins in the environment to create quantum error correction codes~\cite{Hirose16,Cramer16,Taminiau14,Waldherr14}, that could be further tailored to the measured noise spectrum~\cite{Layden18}. An alternative strategy is to optimize dynamical decoupling sequences~\cite{Qi17,Biercuk09,Quiroz13,Uhrig07,Farfurnik15}, for example to allow both noise suppression and quantum sensing~\cite{Poggiali18}. It is then essential to test whether the reconstructed environment model has predictive power. 
\begin{figure}
\includegraphics[width=\columnwidth]{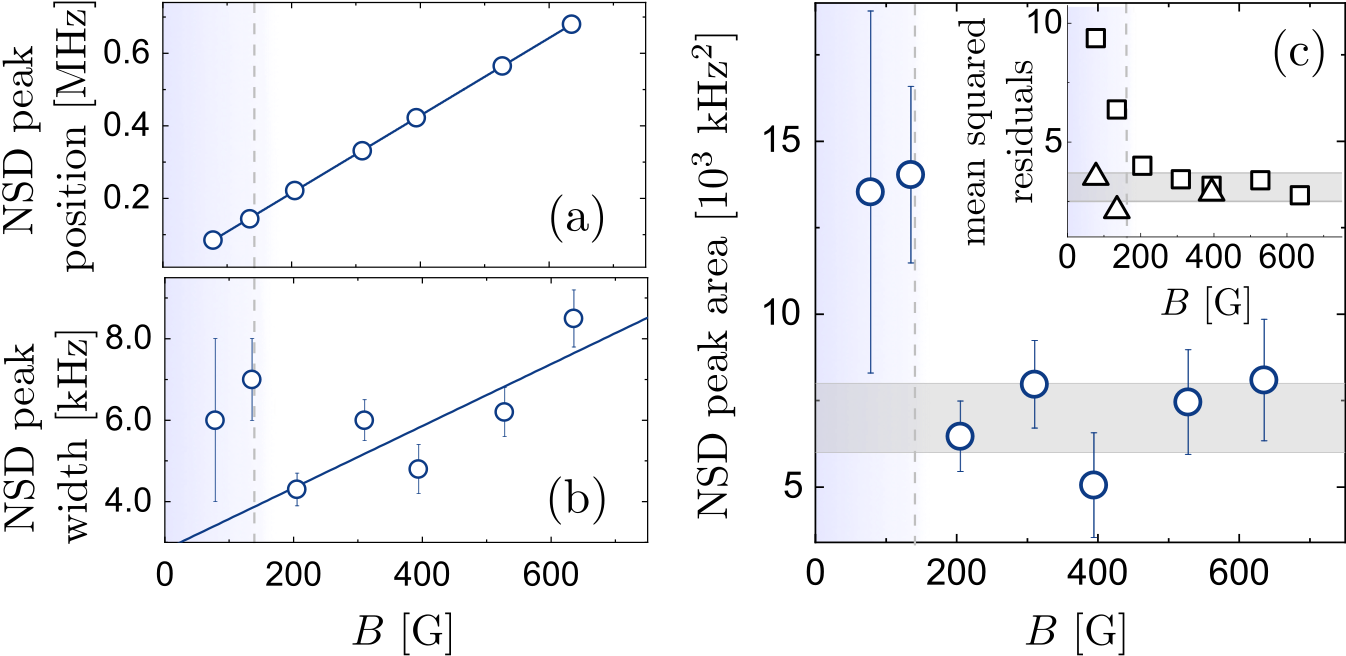}
\caption{NSD peak across the quantum-to-classical spin bath transition. Peak position (a), width (b), and area (c), are obtained from a Gaussian fit of the measured NSD, and reported as a function of the magnetic field strength. The blue line in (a) is a linear fit, consistent with the expected coupling frequency for a $^{13}$C  spin bath. 
The vertical dashed line represents a guide for the eye indicating a sudden change of the bath behavior   -- where the condition $R\gg1$ is no longer fulfilled. The gray horizontal region in (c) denotes the mean value $\bar{A}$ and standard deviation $\sigma$ of the NSD area for $B>150$~G. Values at field $B\leq150$~G deviate from $\bar{A}$ by $>6\,\sigma$. Inset: mean squared residuals of the experimentally observed coherence with the simulation obtained from the measured NSD (squares) and with the 2-model simulation (triangles). Each point results from several datasets collected under different controls~\cite{Suppl}.
}
\label{fig:NSDvsB}
\end{figure}

\section{ Validity and limits of the classical noise model }
Having devised a practical protocol to reconstruct the NV environment, we implement it at different magnetic field intensities, to test whether we can obtain  a predictive model of the spin environment over a range of conditions where either classical or quantum properties of the bath are expected to be visible~\cite{Reinhard12}.
While we expect the spin bath effects on a central spin qubit to be always described by a classic noise source model~\cite{Crow14}, noise spectroscopy  allows us to mark the boundary between quantum and classical regime.
The bias magnetic field applied along the NV spin not only changes the NSD central frequency ($^{13}$C Larmor frequency $\omega_L$), but also its properties. 

In the weak coupling  regime, when $R=\omega_h^\perp/\omega_L\ll1$ for most nuclei, the unpolarized nuclear bath can be described as formed by classical randomly-oriented magnetic dipoles~\cite{Suppl}. The  orthogonal component of each nuclear dipole $\sigma_n^\perp$ undergoes Larmor precession around the external magnetic  field. The coupling to the spin qubit thus assumes the form of an effective dephasing Hamiltonian  $\mathcal{H}=\gamma \beta(t) \sigma_z^{\mbox{\tiny{NV}}}$ with $\beta(t)$ a  time-varying mean field with stochastic amplitude and phase, which can be characterized by its NSD. 
In Fig.~\ref{fig:NSDvsB} we plot peak center, width, and area of the measured $S(\omega)$. The center scales linearly with the magnetic field, with a slope of $1.069(2)$~kHz/G, the gyromagnetic ratio of $^{13}$C (Fig.~\ref{fig:NSDvsB}.a). We can ascribe the small increasing trend of the NSD  width with increasing magnetic field to variations in the internal bath dynamics, as we expect more spin flip-flops at higher fields as the energy of all nuclear spins become dominated by the Zeeman energy and become energetically favorable. A detailed study of this effect, which can be included in the classical bath model, goes however beyond the scope of this work.
Remarkably instead, the NSD width and area (Fig.~\ref{fig:NSDvsB}(b) and Fig.~\ref{fig:NSDvsB}(c)) show a discontinuity at $B\sim 150$~G, where $R\sim1$, indicating a sudden change in the bath properties that we can associate with the boundary between the quantum and classical regimes. 

Having gained in principle a full picture of the NV spin environment -- noise spectrum of the bath and coherent interaction with nearby impurities, we want to confirm this to be a predictive model of the spin evolution under different kinds of time-dependent control, beyond monochromatic filters. We thus use the measured spectrum and hyperfine couplings to simulate the spin coherence under other kind of DD sequences, and we compare this prediction to measurements.

We calculate the residual coherence after a given $n$-pulse sequence, $P_n$, as due to both the spin bath and the $m=3$ observed strongly-coupled single spins,
 \begin{equation}
P_n(T) = \frac{1}{2}\left( 1 + e^{-\chi_n(T)}\prod_{i=1}^m M_n^{(i)}(T) \right).
\end{equation}
Here, $\chi_n(T)$ is obtained from the measured NSD, whereas $M_n(T)$ is extracted by evolving the spin under conditional evolution operators~\cite{Suppl}.
\begin{figure}[thb]
\includegraphics[width=\columnwidth]{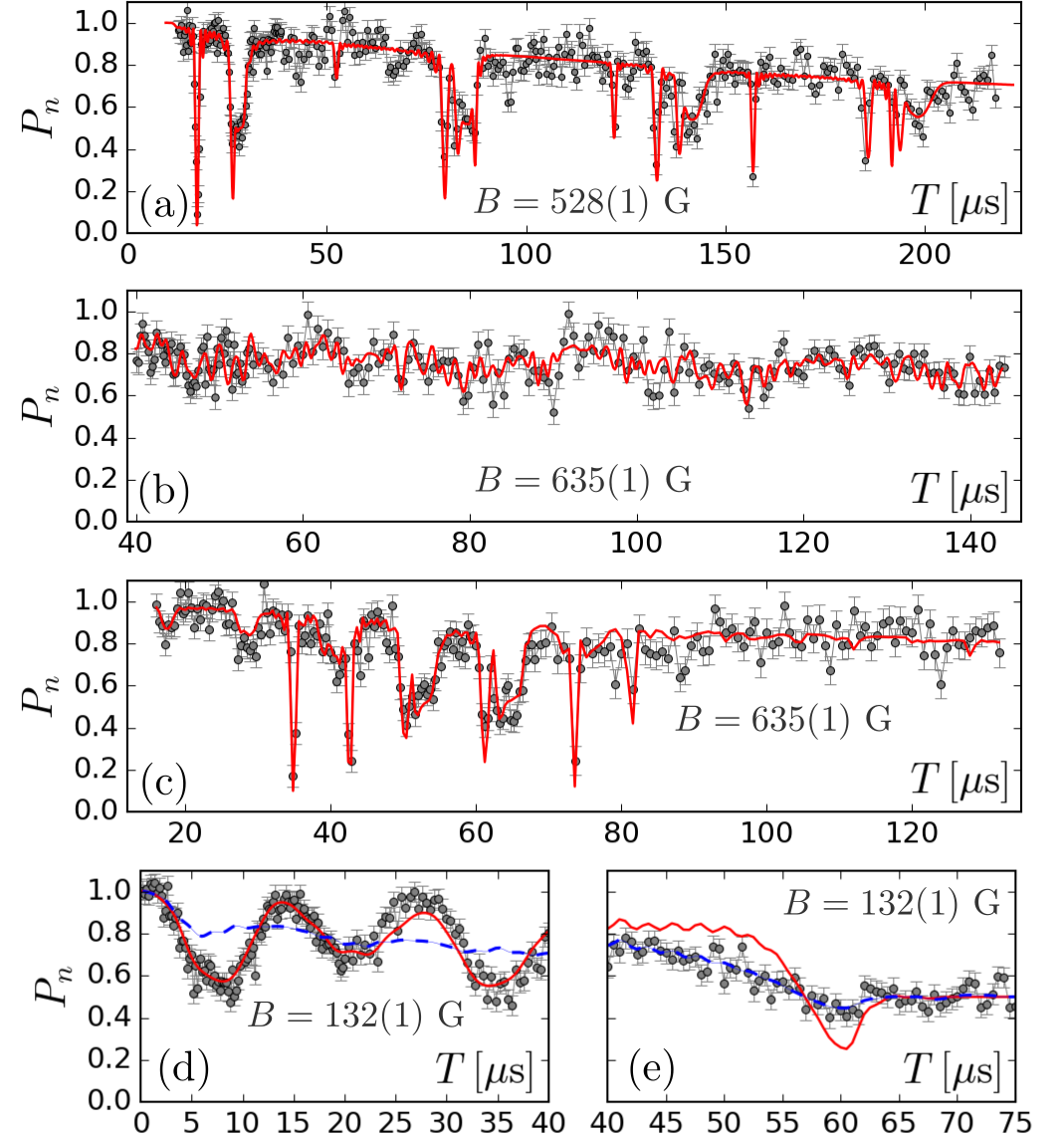}
\caption{\textbf{Time evolution of spin coherence under different DD  sequences.} Dots are experimental data with statistical error. Red lines are the predicted coherence simulated using the measured environment, with no free parameters. (a) Four repetitions of XY-8 ($n=32$). (b) UDD with $n=32$. (c) AXY-8, with $n=40$ pulses at positions $t_{ij}=\frac{T}{80} (10 i+j-8)$, with $i=1...8$ and $j=1...5$
\cite{Suppl}. (d-e) Low-field results for (d) spin-echo and (e) Uhrig $n=32$. The blue dashed line is obtained from the NSD best fit to a set of multipulse sequences (Uhrig, AXY-4, and AXY-8). Mean squared residuals of the experiments with simulation are summarized in the inset of Fig.\ref{fig:NSDvsB}(c).
}\vspace{-12pt}
\label{fig:simCoh}
\end{figure}

Figure~\ref{fig:simCoh} compares simulation and experimental $P_n$. In addition to  equispaced sequences, such as spin echo and XY8-N used for noise spectroscopy, we implemented Uhrig dynamical decoupling (UDD~\cite{Uhrig07}), which was recently used to detect remote nuclear spin pairs~\cite{Zhao11}, as it highly suppresses the effect of  coupling to single nearby nuclei; and adaptive XYN sequences, AXY-N~\cite{Casanova15}, which have been proposed to improve the robustness and discrimination of single nuclear spins.

At high field, the excellent agreement between data and simulations (see also inset of Fig.~\ref{fig:NSDvsB}(c), squares) demonstrates that the spin bath can be described independently of the NV dynamics, since the coupling of the NV center to the spin bath can be neglected compared to the bath internal energy. 
In the strong-coupling regime, at $B\leq150$~G, 
we expect the bath dynamics to be modified by the controlled NV dynamics due to the back action of the NV onto the bath itself.
In other words, one single classical model is no longer suitable to describe the bath when the NV dynamics is driven by different kinds of DD sequences. 
We observe indeed that the NSD measured with equispaced sequences does not predict correctly the measured coherence independently of the applied control.
However, since the NV spin is a simple two-level system, once fixed the control sequence acting on the NV spin we should be able to find a classical model of the spin bath~\cite{Crow14} (we note that we still can correctly model the contribution from the coherently coupled nuclear spins using the same parameters and Hamiltonian as in the high-field regime). 
We find that two classical noise spectra are enough to achieve predictive results (Fig.~\ref{fig:simCoh} (d,e), and inset of Fig.~\ref{fig:NSDvsB}(c)).
The NSD extracted with the  method described in Sec.~\ref{Sec:spectroscopy} correctly describes the coherence for DD sequences with small $n$ numbers (e.g. Hahn echo in Fig.~\ref{fig:simCoh}~(d)). On the other hand, the spin dynamics under sequences with a large number of pulses can be predicted by an alternative NSD line shape, obtained from the simultaneous fit of $P_n$ under different multi-pulse controls, which fails in turn to predict Hahn echo (Fig.~\ref{fig:simCoh} (d,e)). The two line shapes differ from each other significantly~\cite{Suppl}, but this combined two-model picture well describes the spin behavior under all the explored controls (inset of Fig.~\ref{fig:NSDvsB}(c), triangles).

\section{ Conclusions} 
We have experimentally demonstrated a method to spectrally characterize the nuclear spin environment of NV centers, even when the resulting NV coherence time is short. The environment comprised both nearby nuclei, that induce coherent modulations, as well as a larger ensemble of nuclear spins, which we aim to model with a classical bath. 
Our method allows identifying the characteristic parameters of both components of the environment (Hamiltonian of nearby nuclei and NSD of the bath). 
The reconstruction of the full environment model can be then used to predict the NV coherence even when the spin dynamics is driven by different kinds of control. 

In a weak coupling regime, at high magnetic fields, the environment model fully predicts the measured spin coherence under various control sequences. At low magnetic fields, where the quantum dynamics of the nuclear spin bath is expected to have larger influence, still we 
can  always identify a classical noise model describing the NV central spin decoherence, even if decoherence is fundamentally induced by the nuclear spin bath via entanglement with the NV center. However, due to the control-driven qubit back action on the bath, the classical noise model is not generally predictive, and we find that different environment models  are  needed  to describe the evolution under different types of applied controls.  

By studying the validity and limits of a robust environment characterization protocol, able to address a complex quantum environment and provide a simplified (classical) model, our results pave the way to more robust quantum devices, protected by noise-tailored error correction  techniques.

The authors thank F.~S.~Cataliotti for critical reading of the manuscript, and M.~Inguscio for supporting the project. This work was supported by EU-FP7 ERC Starting Q-SEnS2 (Grant n. 337135) and by NSF grant EECS1702716. 

\bibliography{Biblio}

\section{Supplemental material}

\subsection{Experimental system}

The qubit system used in experiments was given by the $m_s=\{0,-1\}$ levels of a single deep nitrogen-vacancy (NV) center in an electronic-grade bulk diamond (Element6), with nitrogen concentration $[^{14}$N$] < 5$ ppb. The spin qubit environment was given by an ensemble of $^{13}$C nuclear spins, with  $1.1\%$ natural abundance. 
In the experiment, the NV electronic spin is addressed and manipulated via optically-detected magnetic resonance, as schematically depicted in Fig.~1 of the main text. A home-built confocal microscope enables the optical initialization of the electronic spin $S = 1$ of a single NV center in the $m_s = 0$ spin projection of the ground state $\ket{\mbox{g}}$. Resonant control is performed by irradiating the spin qubit with microwave radiation delivered by an antenna. More details can be found in \cite{Poggiali17,Poggiali18}.

A  Ramsey free-induction-decay experiment features the presence of at least one strongly-coupled nucleus close to the NV spin (coupling strength $\omega_h/2\pi\sim 700$ kHz), apart from the nitrogen composing the NV center itself. The coupling of the NV spin to the carbon nuclear spin bath, is visible in spin-echo experiments at low magnetic field strength ($B \lesssim 150$~G) as periodic collapses and revivals of coherence spaced by the bare $^{13}$C Larmor period, as expected \cite{Childress06}.

To implement our method for environment spectral reconstruction, we measured the generalized coherence decay time $T_2^L$ under pulsed dynamical decoupling (DD) sequences. We collected data for a large set of $t_1$ values $(0.15-6.5\,\mu $s$)$, and repeated each measurement for different number of pulses, typically multiple of $n=8$. 
We use the XY-$8$ sequence \cite{Maudsley86,Gullion90} as a base cycle. We found that in our setup applying the $\pi/2$ pulses along Y improves coherence with respect to using $\pi/2$ pulses along X, and thus used this choice of phase. Still, we maintained the usual ``XY'' sequence nomenclature  throughout the manuscript. In all the experiments, we also average the signal over two measurements where the last $\pi/2$ rotation is applied respectively along Y and $-$Y, to improve reproducibility of the signal.
For $t_1$ values with coherence showing a fast decay, we also included $n=1, 4$, and $12$. By combining the data sets associated to each $t_1$ value, we separated the contribution of different noise frequency components $\omega=\pi/2t_1$. We then categorized the coherence behavior as a function of $n$ into two different cases: decay and modulation, 
as presented in Fig.2(d) of the main text.

\subsection{Classical model for a quantum bath}
\label{sec:classicalquantum}
While  decoherence is usually described as the loss of quantum coherence due to  the creation of system-environment entanglement and subsequent tracing over of the environment, non-unitary dynamics can also arise from the stochastic variation of a classical parameter\cite{Zurek91,Helm09}. 
If decoherence arises from the stochastic variation of a classical parameter $\{B_k\}$, then the system dynamics can always be written as a random unitary map (RUM).

What is more, it has been shown~\cite{Helm11,Crow14} that for a qubit (and qutrit) it is \textit{always} possible to rewrite any decoherence process as a RUM, even when it arises from an underlying quantum bath. 
Note however that the classical model for a qubit bath is in general non-predictive: although for a given (quantum) bath and qubit (controlled) dynamics we can find a RUM that describes the qubit evolution, the generators of this RUM do not in general describe the qubit evolution when subjected to a different control. For the spin bath considered in our work, we can show that the classical model becomes predictive in the limit of weak coupling between the system and the environment.

The Hamiltonian of the system plus bath is
\begin{equation} \mathcal{H}=\mathcal{H}_S+\mathcal{H}_B+\mathcal{H}_{SB} \label{eq:Ham}
\end{equation} 
with $[\mathcal{H}_S,\mathcal{H}_{SB}]=0$ for  a generalized dephasing noise:
\[\mathcal{H}_{SB}=\sum_a \upsilon_a(t)\ket{a}\!\bra{a}\otimes B_a.\]
Here $\upsilon_a(t)$ is a scalar representing the coupling to  level $a$  of the quantum system,  which can be time dependent if pulsed control is applied on the system (we restrict here to bang-bang control).
As $[\mathcal{H}_S,\mathcal{H}_{SB}]=0$, we can set $\mathcal{H}_S=0$ without loss of generality. Note that for a classical stochastic noise $\mathcal{H}_B=0$ and $B_a$ are stochastic (time-dependent) scalar functions. 

The total system evolves in time as $\rho_{SB}(t)=U(\rho_S(0)\otimes \rho_B(0))U^\dag$. Assuming we prepare the system in a superposition state $\psi_S(0)=\frac1{\sqrt2}(\ket{a}+\ket{b})$, we want to measure the survival probability 
$P(T)=\bra{\psi_S(0)}\textrm{Tr}_B[\rho_{SB}(T)]\ket{\psi_S(0)}$. Since we can write the propagator as $U=\sum_a \ket{a}\!\bra{a}U_a$, given the form of the Hamiltonian, the probability can be simplified to (here assuming a qubit system):
\begin{equation}
P(T)=\frac12\left(1+\textrm{Re}[\textrm{Tr}_B(U_0\rho_BU_1^\dag)]\right)
\label{eq:Psurvival}	
\end{equation}

Note that for a classical bath the noise dependence on $a$ is encoded in $\upsilon_a$ and the $U_a$ are scalars,  thus we need to replace the trace over the bath with an average. We then have
\begin{equation}
\begin{array}{l}
P(T)=\frac12\left(1+\langle e^{-i\int_0^T [\upsilon_a(t)+\upsilon_b(t)]B(t)dt} \rangle\right)\\
\approx\frac12\left(1+ e^{-\frac12\int_0^T [\upsilon_a(t)+\upsilon_b(t)][\upsilon_a(t')+\upsilon_b(t')]\langle B(t)B(t')\rangle dt dt'} \right),
\end{array}
\label{eq:Pclassical}	
\end{equation}
where the approximation is exactly valid for a Gaussian stationary noise. The last term can be rewritten in terms of the overlap between the DD sequence filter and the noise spectral density (the Fourier Transform of the correlation $\langle B(t)B(t')\rangle$). 

For a quantum bath, it is convenient to first remove the bath internal Hamiltonian by going into the interaction picture. We thus obtain
\begin{equation} B(t)=e^{i\mathcal{H}_Bt}Be^{-i\mathcal{H}_Bt} \label{eq:Hamint}\end{equation} 

Then the evolution is governed by the propagator:
\begin{equation} \begin{array}{l}
U(T)=\mathcal T \exp\left[-i\int_0^T\sum_a \upsilon_a\ket{a}\!\bra{a} \otimes B_a(t)dt\right] 
\\
\approx \exp\left[-i\sum_a \ket{a}\!\bra{a}(\overline B_a+\overline C_a+\dots)T\right],
\end{array}
\label{eq:Ut}
\end{equation} 
where the expansion is given by the Magnus expansion with the first two orders given by:
\begin{equation}
\overline B_a=\frac{1}{T}\int_0^Tdt \upsilon_a(t)B_a(t) 
\end{equation}\begin{equation}
\overline C_a(T)\!=\!\frac{-i}{2T}\!\int_0^T\!\! dt_1\!\!\int_0^{t_1}\!\!\!dt_2\,\upsilon_a(t_1)\upsilon_a(t_2)[B_a(t_1), B_a(t_2)]
\label{eq:AHT}
\end{equation}

We consider a bath of non-interacting spins that interact with the quantum system as
\begin{equation}
B_a=\frac{1}2\sum_k \omega_h^{k,\parallel}\sigma_z^k+\omega_h^{k,\perp}\sigma_x^k
\label{eq:SpinBathNV}
\end{equation}
and have internal Hamiltonian
\begin{equation}
\mathcal{H}_{B_s}=\frac{\omega_0}2\sum \sigma_z^k.
\label{eq:SpinBathHamNV}
\end{equation}
 In the interaction picture of the bath internal Hamiltonian, the noise operator  becomes
\begin{equation}
B_a(t)=\frac{1}2\sum_k \omega_h^{k,\parallel}\sigma_z^k\cos\theta_k+\omega_h^{k,\perp}[\cos(\omega_0 t)\sigma_x^k+\sin(\omega_0 t)\sigma_y^k].
\label{eq:SpinBathNVInt}
\end{equation}

The survival probability can be written as
\begin{equation}
P(T)=\frac12\left(1+\prod_k \textrm{Tr}\left[e^{i(\overline B^k_b+\overline C^k_b)}\rho^k_Be^{-i(\overline B^k_a+\overline C^k_a)}\right]\right)
\label{eq:Pquantum}	
\end{equation}
We can simplify this further by assuming $\rho_B=\openone/N$ and considering the system to be a qubit (two-level system).
For a qubit, the leading contribution is given by $\overline B^k$, while the terms $\overline C_k$ cancel out,  if we neglect the third order term $\propto[\overline B,\overline C]$, as $\overline B_0=\overline B_1$ and  $\overline C_0=-\overline C_1$. Given $\frac12\textrm{Tr}\left[e^A\right]=\cosh(\|A\|_1/2)$ (for traceless operators $A$), we then have
\begin{equation}
\prod_k\frac12\textrm{Tr}\left[e^{-2i\overline B^k}\right]=\prod_k\cosh(\|-2i\overline B^k\|_1/2)\approx e^{-2\sum_k \|\overline B^k\|_1^2}
\label{eq:spinRamsey}
\end{equation}
We finally obtain the residual coherence for a Ramsey (free evolution)
\begin{equation}
P(T)=\frac12+\frac12e^{-\frac{T^2}2\sum_k |\omega_h^{k,\parallel}|^2}\!\!\exp\left[-2\sum_k  |\omega_h^{k,\perp}|^2\frac{\sin(\omega_0T/2)^2}{\omega_0^2}\right]	
\end{equation}
and for a DD sequence such as CPMG with $n$ cycles:
\begin{equation}
P(T)=\frac12+\frac12\exp\left[-2\sum_k |\omega_h^{k,\perp}|^2 \frac{ \sin ^4\left(\frac{T \omega_0}{4}\right) \sin ^2\left(n T \omega_0\right)}{\omega_0^2 \cos ^2\left(\frac{T \omega_0}{2}\right)}
\right]
\label{eq:spinDD}
\end{equation}

These results can be modeled by assuming the presence of a classical magnetic field, with a static component with rms $\Delta B^2=\sum_k |\omega_h^{k,\parallel}|^2$ (which gets refocused by DD sequences) and a stochastic one, with spectrum $S(\omega)= \sum_k  |\omega_h^{k,\perp}|^2\delta(\omega-\omega_0)$. Since the $\overline B$ terms only include a simple average over the DD sequence, if higher terms can be neglected, this classical model of the spin bath has predictive powers, for any other pulsed DD sequence. 
The condition for neglecting the higher orders is related to the ratio $\|\omega_h\|/\omega_0$ (or, in the case of DD sequences, $\omega_h^{\perp}/\omega_0$). Here, bath couplings without an index indicate the ``typical'' interaction strength (indeed, a discrete number of more strongly coupled bath spins could be treated separately, following their coherent evolution, as done in the main text).
For strong noise ($\|\omega_h\|\approx\omega_0$), where one needs to take into account higher order terms, it would still be possible to find a classical model but the model must be varied every time one select a different control scheme, even when we limit the control to $\pi$-pulses (DD sequences).


\begin{table}[t]\renewcommand{\arraystretch}{1.2}
\caption{Comparison between three different simulated NSD and their correspondent reconstruction from the simulated data. The NSD is defined by its center $\nu_L=\omega_L/2\pi$, amplitude $A$, width $\sigma$ and offset $y_0$. The original parameters for $B=635(1)$~G correspond to the dashed red line in Fig.~3(d) of the main text, and the parameters of the reconstructed NSD using the first two harmonics correspond to the blue line in the same figure of the main text.}
\begin{tabular}{cccccc}
&&\multicolumn{1}{c}{} & \multirow{ 2}{*}{NSD}& \multicolumn{2}{c}{Reconstructed NSD} \\ \cline{5-6}
&&\multicolumn{1}{c}{} &  & $\;\;0^{th}$-order$\;\;$ & $\;\;1^{st}\&2^{nd}$-orders$\;\;$ \\ \cline{3-6}
\parbox[t]{2mm}{\multirow{3}{*}{\rotatebox[origin=c]{90}{$B=208(1)$~G}}}&\phantom{...}&$\nu_L\,^\dag$ &  221.2 & 220.8(1) & 221.202(3) \\
&&$A\,^\dag$ & 600  & 297(5) & 600.9(5)  \\
&&$\sigma\,^\dag$ & 4.3 & 9.9(2) & 4.304(3)   \\
&&$y_0\,^\dag$ & 5.6 & 6.23(4) & 6.791(4)  \\
\cline{3-6}
\end{tabular}
\begin{tabular}{cccccc}
&&\multicolumn{1}{c}{} & \multirow{ 2}{*}{NSD}& \multicolumn{2}{c}{Reconstructed NSD} \\ \cline{5-6}
&&\multicolumn{1}{c}{} &  & $\;\;0^{th}$-order$\;\;$ & $\;\;1^{st}\&2^{nd}$-orders$\;\;$ \\ \cline{3-6}
\parbox[t]{2mm}{\multirow{3}{*}{\rotatebox[origin=c]{90}{$B=635(1)$~G}}}&\phantom{...}&$\nu_L\,^\dag$ & 679.9 & 680(2) & 679.91(4)  \\
&&$A\,^\dag$ & 380  & 197(32) & 381(2)  \\
&&$\sigma\,^\dag$ & 8.5 & 16(3) & 8.70(4)   \\
&&$y_0\,^\dag$ & 3.7 & 3.66(4) & 4.337(2)  \\
\cline{3-6}
\end{tabular}
\begin{tabular}{ccccccc}
&&\multicolumn{1}{c}{} & \multirow{ 2}{*}{NSD}& \multicolumn{3}{c}{Reconstructed NSD}\\ \cline{5-7}
&&\multicolumn{1}{c}{} &  & $\;\;0^{th}$-order$\;\;$ & $\;\;1^{st}\&2^{nd}$-orders$\;\;$ & $\;\;1^{st}$--$10^{th}$-orders$\;\;$ \\ \cline{3-7}
\parbox[t]{2mm}{\multirow{3}{*}{\rotatebox[origin=c]{90}{$B=700$~G}}}&\phantom{...}&$\nu_L\,^\dag$ & 750 & 748(2) & 750.0(2) & 750.02(2) \\
&&$A\,^\dag$ & 600  & 266(34) & 601(13) & 603(1) \\
&&$\sigma\,^\dag$ & 9 & 21(4) & 9.0(1) & 9.01(2) \\
&&$y_0\,^\dag$ & 5 & 5.00(5) & 5.860(4) & 5.983(2) \\
\cline{3-7}
&&\multicolumn{5}{l}{$^\dag$ Values have units of kHz.}\\
\end{tabular}
\label{tab:0orderFit}
\end{table}
\subsection{Noise spectral density analysis}
\label{Sec:NSDanalysis}
When we characterized the noise spectral density (NSD) using equidistant-pulse sequences, we found a discrepancy between the results obtained for the first collapse, i.e. the $0^{th}$-order harmonic, and the higher order collapses.
The problem stems from the very fast  decay in terms of the number of pulses for the first collapse. This means that using large number of pulses to extract $T_2^L$ is not possible. For low number of pulses, unless one performs a deconvolution of the sinc-like modulation, the filter induces an additional broadening of the NSD. An accompanying issue is that XY-$8$ sequences can no longer be used for the measurement of $T_2^L$, but one needs to use alternative equidistant-pulse sequence, as CPMG or XY-$4$, which although robust enough for small $n$, might still introduce some pulse error effects.

As explained in the main text, to overcome this issue, we used higher harmonics of the filter function in order to reconstruct the NSD. To confirm our experimental results and validate our noise spectroscopy method we simulated the decay of a spin qubit under DD sequences in the presence of a simple noise model. By comparing the experimental and simulated results we verified self-consistency of the method. We considered a Gaussian noise with NSD
\begin{equation}
S(\nu)=y_0+A\,e^{-(\nu-\nu_L)^2/(2\sigma^2)},
\end{equation}
where $\nu_L= \omega_L/(2\pi)$ is the Larmor frequency of a carbon nuclear spin bath. We then calculated the temporal dependence of the spin coherence under control sequences with different numbers of equidistant pulses, and we used this coherence to reconstruct the simulated NSD with the same procedure used to treat the experimental data. 

As shown in Fig.~3(d) of the main text,  the $0^{th}$~order peak of the reconstructed spectrum from the simulated data shows the same trend of the experimental data (Fig.~3(b)), and exhibits highly significant disagreement from the original spectrum.
Using instead the $1^{st}$ and $2^{nd}$ harmonics we reliably reconstruct the NSD peak (see Table~\ref{tab:0orderFit}).
From these results we can observe how the original NSD is not correctly reconstructed from the $0^{th}$-order harmonic, meanwhile the first two harmonics are enough to obtain the NSD with good precision.
We also performed the same analysis taking the first ten harmonics, with similar results. 
For the reconstruction, we simulated the coherence only for $n\geq8$, including the points with $n<8$ increases the percentage error on the estimated amplitude and width by a factor of 100 and 40, respectively. The estimation of the center was the same, within the error bars, in all cases.

\begin{figure}
\includegraphics[width=\columnwidth]{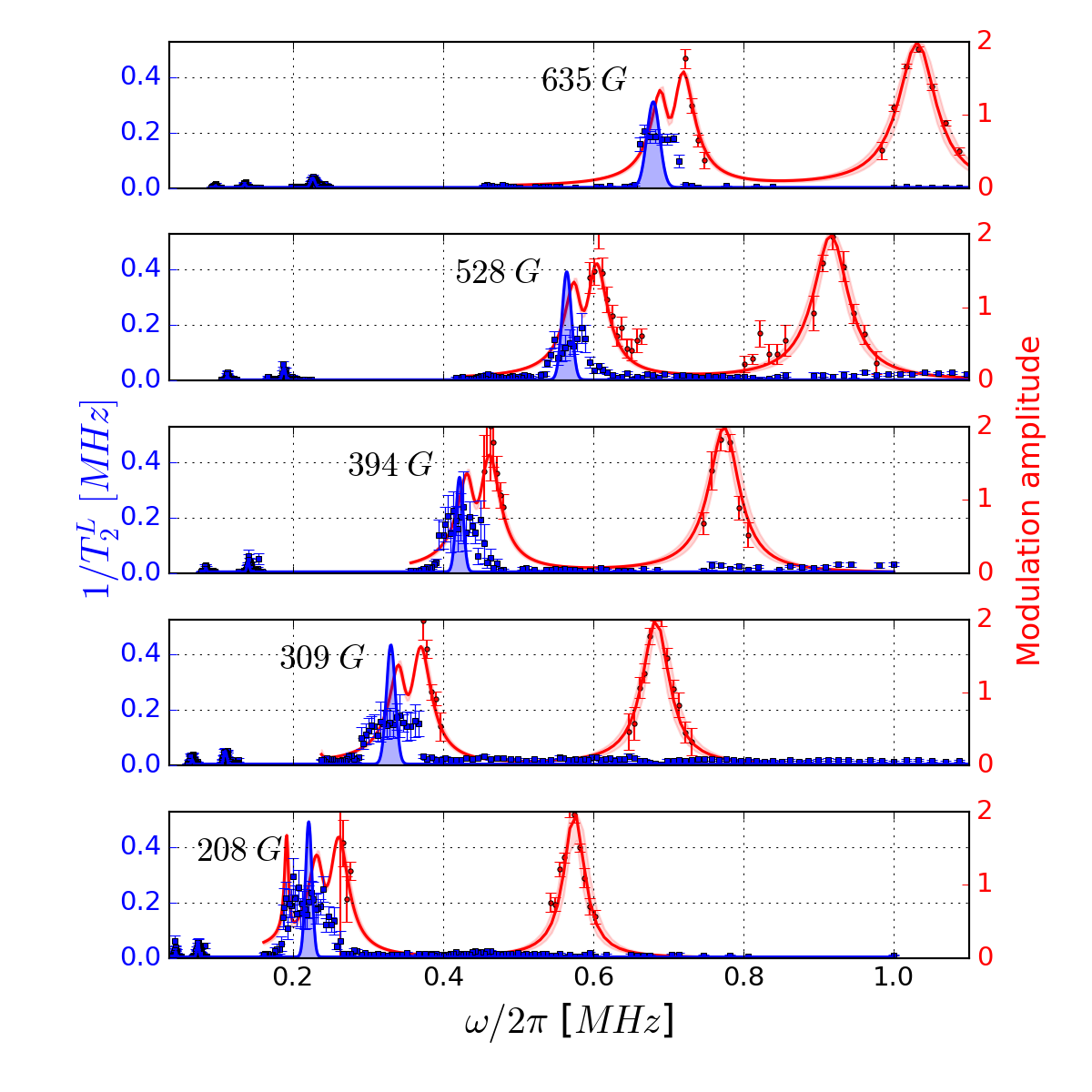}
\caption{NSD and coupling with nearby nuclei. We report $1/T_2^L$ (blue, left-hand side vertical scale) and the amplitude of the observed coherent modulations (red, right-hand side vertical scale), for different magnetic field strengths ($B=208-635$~G). Blue lines are the fit of high-order harmonics to extract the NSD, and red lines are the simulations of the completely resolved nearby Carbons. The red shadow describes the uncertainty on the estimation of the coupling strength components. }
\label{fig:coherenceNall}
\end{figure}

In experiment, we measured the NSD at different magnetic fields. The parameters that describe the NSD  are listed on Table~\ref{tab:expResults}, they correspond to a Gaussian distribution $S(\omega)=y_0+A\,e^{-(\omega-\omega_L)^2/(2\sigma_\omega^2)}$. Figure~\ref{fig:coherenceNall} completes the dataset shown in Fig.~2 of the main text.
\begin{table*}\renewcommand{\arraystretch}{1.2}
\caption{Experimental results. Parameters obtained for the NSD peak using only the 0-th order collapse and using higher harmonics.}
\begin{tabular}{|c|cccc|cccc|}
\hline 
\multirow{2}{*}{B-field [G]} & \multicolumn{4}{c|}{$0^{th}$-order} & \multicolumn{4}{c|}{Higher orders: $1^{st}$ \& $2^{nd}$} \\ 
\cline{2-9}
& $y_0^\dag$ & $A^\dag$ & $\nu_L^\dag$ & $\sigma^\dag$ & $y_0^\dag$ & $A^\dag$ & $\nu_L^\dag$ & $\sigma^\dag$ \\ 
\hline 
$635$*  & $0.0191(7)$ & $0.35(8)$ & $0.683(3)$ & $0.018(2)$ & $0.0037(2)$ & $0.38(5)$ & $0.6799(8)$ & $0.0085(7)$ \\
$528$ & $0.0174(8)$ & $0.16(3)$ & $0.571(2)$ & $0.019(2)$ & $0.0040(3)$ & $0.48(7)$ & $0.5647(7)$ & $0.0062(6)$ \\
$394$   & $0.0144(7)$ & $0.25(4)$ & $0.425(2)$ & $0.017(1)$ & $0.0061(3)$ & $0.42(6)$ & $0.4220(6)$ & $0.0048(5)$ \\
$309$   & $0.0260(8)$ & $0.21(3)$ & $0.331(2)$ & $0.018(1)$ & $0.0070(6)$ & $0.53(7)$ & $0.3306(6)$ & $0.0059(5)$ \\
$208$   & $0.016(1)$ & $0.29(2)$ & $0.223(1)$ & $0.0187(8)$ & $0.0056(4)$ & $0.60(8)$ & $0.2212(4)$ & $0.0043(4)$ \\
\hline
\multicolumn{9}{l}{* For 635 G we also used the 3rd order harmonic.} \\ 
\multicolumn{9}{l}{$^\dag$ Values have units of MHz.}\\
\end{tabular} 
\label{tab:expResults}
\end{table*}

\subsection{Spin evolution with coupling to a single nucleus}

The time evolution of the qubit coherence under the effect of a single nuclear spin can be calculated as \cite{Yang08b}
\begin{equation}\label{eq:M}
M(T) = \text{Tr}\left( U_0  U_1^\dag \right)
\end{equation}
where 
$U_0=\cdots e^{-iH_0t_3} e^{-iH_1t_2} e^{-iH_0t_1}$, and $U_1=\cdots e^{-iH_1t_3} e^{-iH_0t_2} e^{-iH_1t_1}$. Here 
$t_i$ is the free evolution time between two consecutive microwave pulses. $H_{0(1)}$ is the Hamiltonian that describes the interaction between the $m_s=0(\pm1)$ electron spin and a nearby nucleus~\cite{Taminiau12}:
\begin{align}
H_0 & = \omega_L \sigma_z, \\
H_1 & = (m_s\,\omega_h^\parallel + \omega_L)\sigma_z + m_s\,\omega_h^\perp \sigma_x,
\end{align}
where $\sigma_i$ are the Pauli matrices on the nuclear spin basis and $m_s$ is the eigenvalue of $S_z$ for the electronic spin.
This treatment allows us to simulate $M(T)$ under different kinds of DD sequences. 

\begin{figure}
\includegraphics[width=\columnwidth]{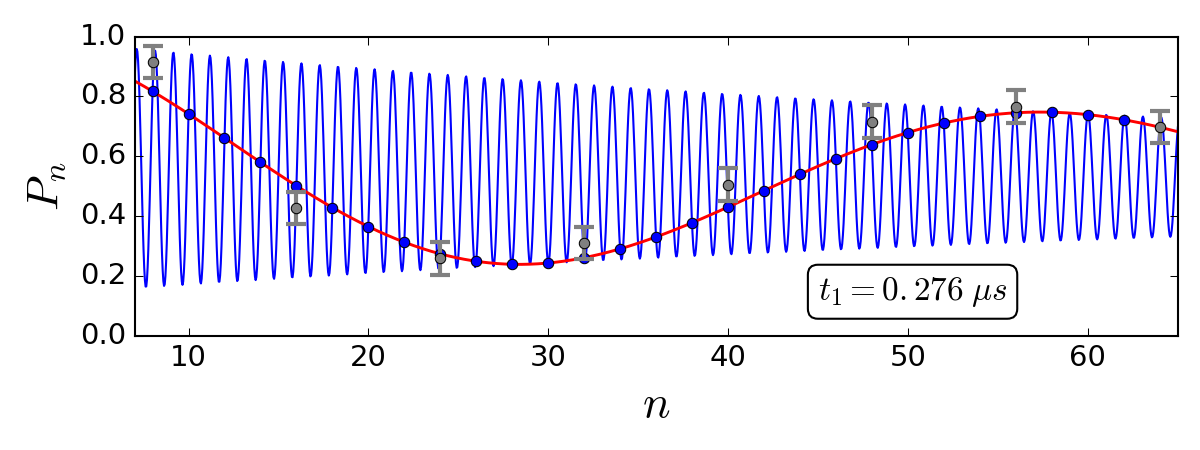}
\caption{
Probability $P_n$ of the electronic spin to be in the state $\ket{-1}$, in terms of the number of $\pi$-pulses $n$ (external bias field $B=(528\pm1)$~G). The gray points are the experimental data, the blue and red solid lines are obtained from fit when considering $\phi$ and $\phi^\prime$, respectively (see text). The blue points show the intersection points of the two curves with even $n$ values.}
\label{fig:coherenceMod}
\end{figure}

To experimentally determine the components of the coupling strength, $\omega_h^\parallel$ and $\omega_h^\perp$, we used Eq.~(5) of the main text together with an analytic expression for $M(T)$ valid  for an even number of equidistant pulses $n$
\begin{equation}\label{eq:M_equi}
M_n(2nt_1) \! = \! 1 \! - \! 2 \! \left( \! \frac{\omega_h^\perp}{\omega_1} \!\right)^2 \sin^2 \! \left( \!\frac{\omega_1 t_1}{2} \! \right) \sin^2 \! \left( \! \frac{\omega_L t_1}{2} \! \right) \frac{\sin^2 \! \left(\frac{n \phi}{2}\right)}{\sin^2 \! \left(\frac{\phi}{2}\right)},
\end{equation}
where $\omega_1 = \sqrt{(m_s\,\omega_h^\parallel + \omega_L)^2 + (\omega_h^\perp)^2}$ is the frequency seen by the spin in the $m_s=\pm1$ state, and the phase $\phi$ is the modulation frequency of the transition probability as a function of $n$, given by
\begin{equation}
\cos \phi = \frac{m_s\,\omega_h^\parallel + \omega_L}{\omega_1} \sin ( \omega_1 t_1 ) \sin ( \omega_L t_1 ) - \cos ( \omega_1 t_1 ) \cos \left( \omega_L t_1 \right).
\label{eq:cosphi}
\end{equation}
We  observe that $\phi$ is small for times $t_1$ corresponding to the minimum values of $M$.

Note that equations~\eqref{eq:M_equi} and \eqref{eq:cosphi} were obtained from Refs.~\cite{Kolkowitz12l,Taminiau12} with the simple transformation $\phi \rightarrow \phi' = \pi - \phi$. 
Both of these conventions are consistent with the model and suitable to evaluate the two components of the hyperfine strength, because they assume the same $M$ values for even integer $n$; nonetheless, they result in a very different periodicity of $P_n$ as a function of $n$, as shown in Fig.~\ref{fig:coherenceMod}, making the analysis of the data very different in terms of the convergence of the fit. 

\begin{figure}
\includegraphics[width=\columnwidth]{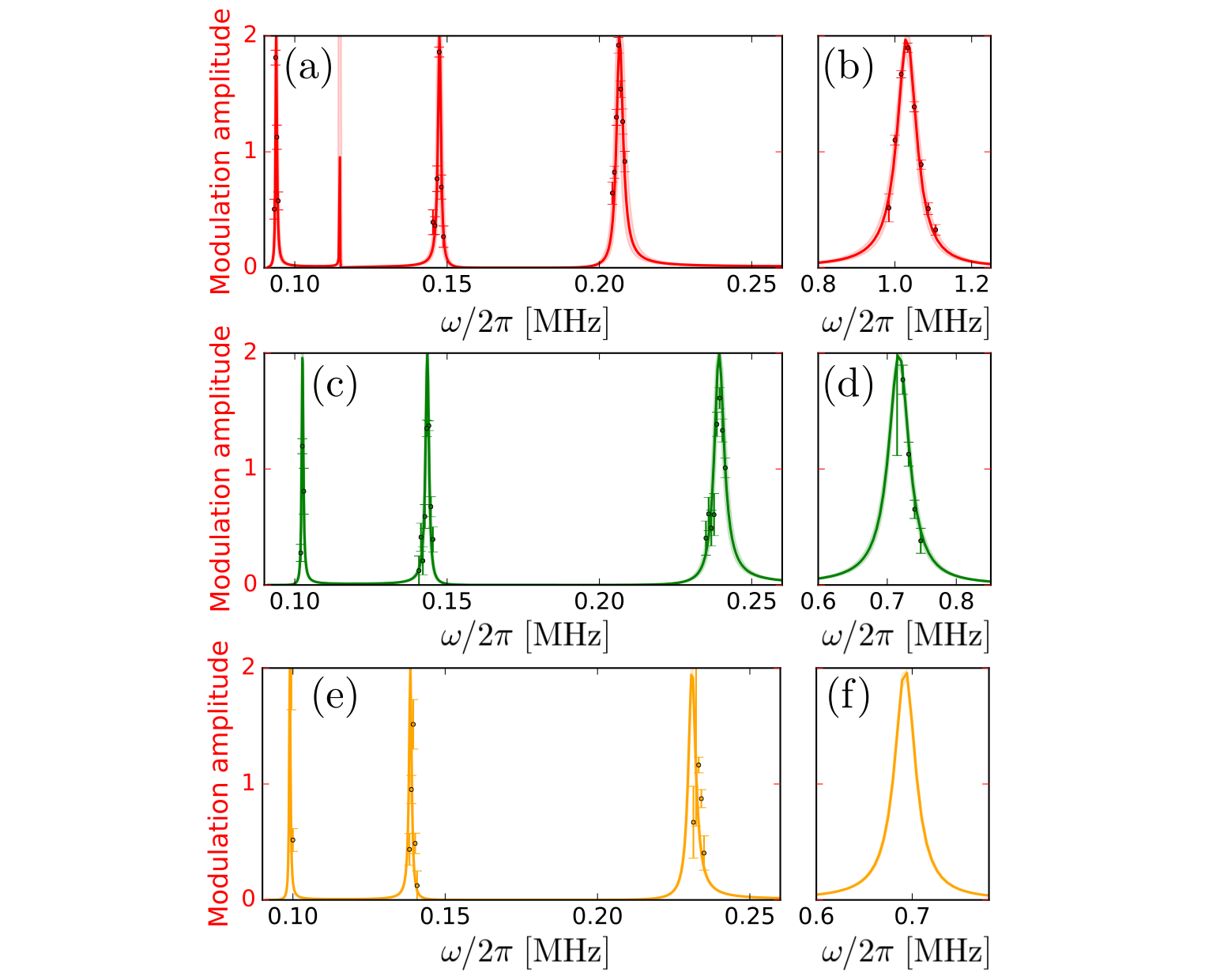}
\caption{
Amplitude of the observed coherent modulations, at $B=(635\pm1)$~G. The points represent the experimental data and the lines represent the simulation after the characterization of $\omega_h^\parallel$ and $\omega_h^\perp$, using Eq.~\eqref{eq:M_equi}. 
The colors red, green and orange represent the different nearby carbons, following the same order than the Table~\ref{tab:coupStr}. Panels (b), (d) and (f) show the modulation for the $0^{th}$-order harmonic of the filter function; (a), (c) and (e) show the modulation amplitudes for higher harmonics of the filter function. Note that the third carbon is so weakly coupled to the NV spin, that the $0^{th}$-order is not distinguishable from the bath for all the investigated magnetic fields (see Fig.~\ref{fig:coherenceNall}).}
\label{fig:modulAmplitud}
\end{figure}

The components of the coupling strength associated with each of the three carbons are shown on Table~\ref{tab:coupStr}. An example of the amplitude of the modulations, used to extract the coupling strengths, is shown in Fig.~\ref{fig:modulAmplitud}.
Two of the three observed nuclei are so weakly coupled that they are not visible in usual Ramsey experiments. The more strongly coupled nucleus is visible with a Ramsey experiment, which still  gives a consistent but less precise estimate of the coupling strength, due to decoherence.

\begin{table}
\caption{Coupling components of the NV spin to three $^{13}$C nuclei.}
\begin{tabular}{cc}
$\omega_h^\parallel /2\pi$ [kHz]& \qquad $\omega_h^\perp /2\pi$  [kHz]\\
 \hline
$-698\pm8$ & $148\pm13$ \\
$-73\pm4$ & $59\pm3$ \\
$-25\pm2$ & $42\pm1$
\end{tabular}
\label{tab:coupStr}
\end{table}

\subsection{Summary of simulation and experiment results}
The predictive capabilities of the measured NSD and the coupling strengths of the nearby Carbons is critical to prove the noise characterization method. We use the experimentally determined environment model to simulate the time evolution of the NV coherence under different control sequences, and we compare the simulation with experiments. In the following, we add detail on the examples shown in the main text, and we present additional data, with other control sequences, which we describe below, to further validate our results. 

\subsubsection{Adaptive XY-8 sequence.}
Nonuniformly-spaced $\pi$-pulse sequences~\cite{Zhao14,Ma15} have been designed to improve resolution of single nuclear spin spectroscopy. In addition, adaptive XY-N (AXY-N) sequences~\cite{Casanova15} have been proposed to enhance robustness by sending the pulses along different axes within each other. AXY-8 
is constructed by eight blocks, each containing five equidistant pulses with specific phases -- known as Knill pulse~\cite{Souza11KDD,Casanova15}. 
The phase distribution of the pulses in each block is:
\begin{align}
\Pi_\phi = \pi_{\phi+\pi/6} \; \-- \; \pi_{\phi}\; \-- \; \pi_{\phi+\pi/2}\; \-- \; \pi_{\phi}\; \-- \; \pi_{\phi+\pi/6}
\end{align}
to form the AXY-8 sequence, the eight blocks must be:
\begin{align}
\Pi_X \; \-- \; \Pi_Y \; \-- \; \Pi_X \; \-- \; \Pi_Y \; \-- \; \Pi_Y \; \-- \; \Pi_X \; \-- \; \Pi_Y \; \-- \; \Pi_X
\end{align}

\begin{figure}
\includegraphics[width=.8\columnwidth]{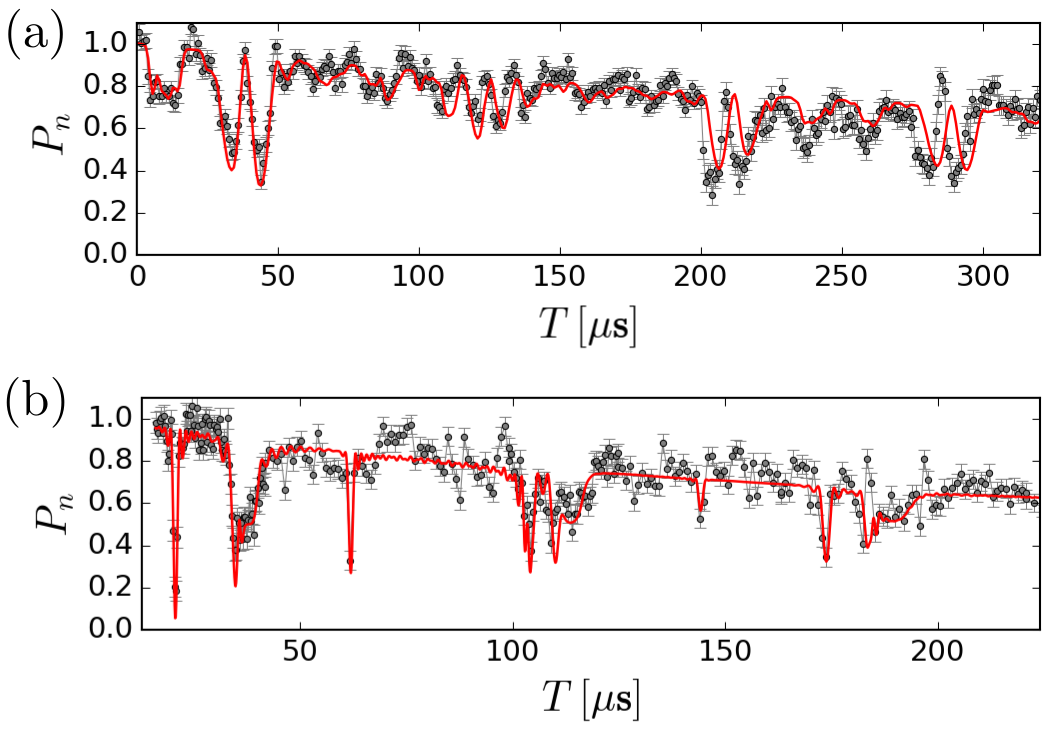}
\caption{Time evolution of spin coherence at field $B=(394\pm1)$~G. The dots are the experimental data, and the red line is the simulation, in both cases the prediction was done using the same environment, \textit{i.e.} same NSD and same nearby nuclei coupling strengths. (a) Two level system formed by $\{\ket{0},\ket{+1}\}$. Sequence: one repetition of XY-4 ($n=4$). (b) Two level system formed by $\{\ket{0},\ket{-1}\}$. Sequence: four repetitions of XY-8 ($n=32$).}
\label{fig:plusminus1}
\end{figure}

\begin{figure*}
\includegraphics[width=.7\textwidth]{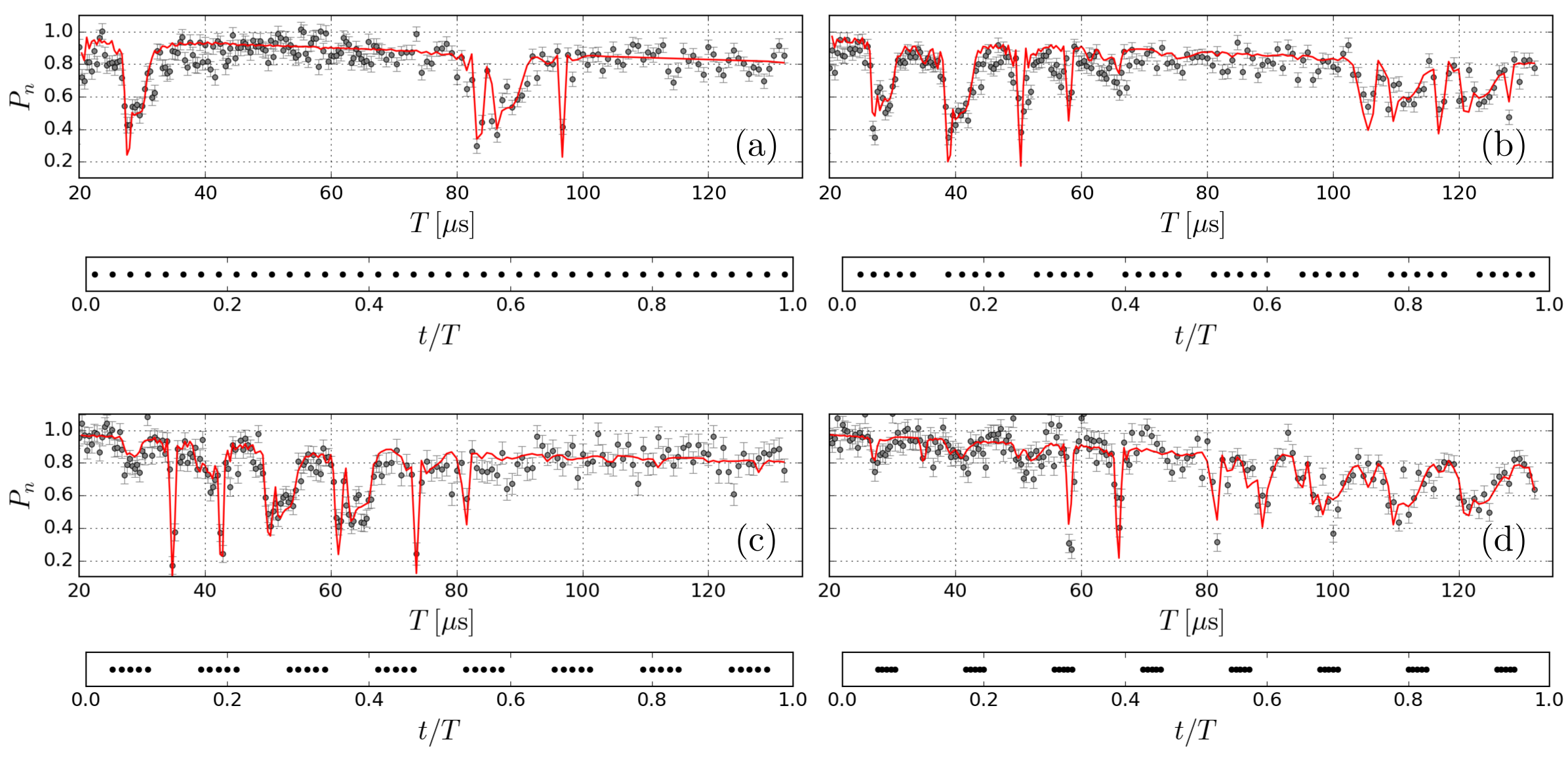}
\caption{Time evolution of spin coherence for different cases of AXY-8 sequences at field $B=(635\pm1)$~G. The gray dots are the experimental data, and the red line is the simulation. Below each plot we show the temporal distribution of the $\pi$ pulses. (a) $r_m=1.0$; (b) $r_m=0.75$; (c) $r_m=0.5$; (d) $r_m=0.25$.}
\label{fig:axy8all}
\end{figure*}

The blocks are equidistant, thus the position in time for each $\pi$-pulse is given by $t_{i,j}(r_m)=(1/\mathcal{N})\left(\frac{2i-1}{2} + r_m\frac{2j-\mathcal{M}-1}{2\mathcal{M}}\right)$ where $\mathcal{M}=5$ is the number of pulses inside each block and $\mathcal{N}=8$ is the number of blocks, $i\in\{1,\dots,\mathcal{N}\}$, $j\in\{1,\dots,\mathcal{M}\}$, and $r_m$ defines how close within each other are the pulses inside each block. With $r_m=1$ the $40$ pulses are equidistant, whereas the limit of $r_m=0$ corresponds to eight $\pi$~pulses (each one formed by the superposition of five pulses). In Figure~\ref{fig:axy8all} we show the case for $r_m=1.0,\,0.75,\,0.5$, and $0.25$. We verified that, as expected, AXY-$8$ and XY-$8$ are more robust compared to CPMG and XY-4 against detuning and pulse shape imperfections, which our simulations do not take into account.

\begin{figure*}
\includegraphics[width=.8\textwidth]{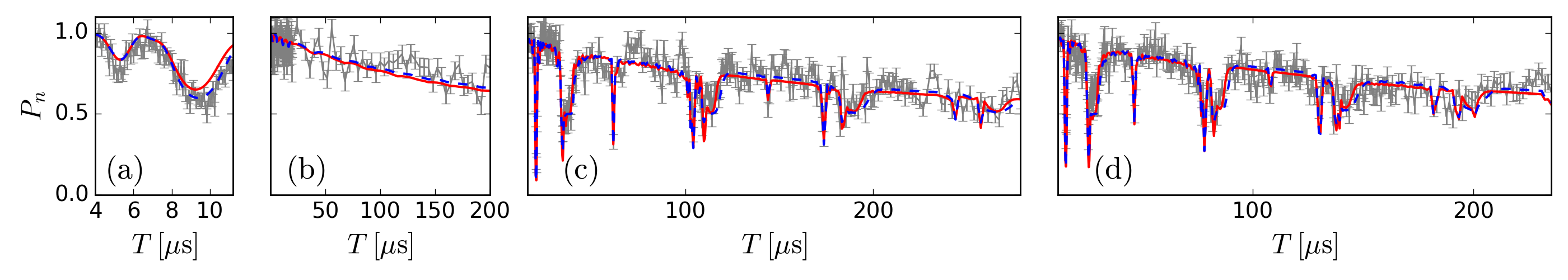}
\caption{Evolution of spin coherence for an external magnetic field $B=(394\pm1)$~G. The dots are the experimental data, the red line is the simulation done with the NSD obtained using the $T_2^L$ method, and the blue line is the simulation done with the NSD obtained by directly fitting the data (see text).  
(a) XY-8 for $n=8$ with $91$ points.
(b) Hahn-echo with $145$ points. 
(c) XY-8 for $n=32$ with $326$ points. 
(d) XY-8 for $n=24$ with $326$ points. For this field we used $888$ points to calculate the mean-squared-residuals.
}
\label{fig:simDat_redChiSqua}
\end{figure*}

\subsubsection{Prediction for the \texorpdfstring{$m_s=+1$}{ms=+1} projection.}  
We can exploit  the spin triplet nature of the NV center  to further validate the fact that the reconstructed environment model is predicted independently of the qubit properties, that is, the environment model extracted from the dynamics of the the two spin states $\ket{0}$ and $\ket{-1}$ can predict also the dynamics of the $m_s=\{0,+1\}$ manifold and not only the $\{0,-1\}$ manifold as presented in the main text.
Figure~\ref{fig:plusminus1} shows two different sequences for the two different spin projections together with their simulation. In both cases we found a good agreement between the prediction and the experimental data.

\subsubsection{Residuals -- An estimation of the prediction capability.}  
The predictive power of the model can be estimated quantitatively by calculating the mean-squared-residuals (defined as the reduced chi squared) between simulation and experimental data:
\begin{equation}
\chi_{\{N-1\}}^2 = \frac{1}{N-1} \sum_{i=1}^N \frac{(s_i-y_i)^2}{\delta_i^2}
\label{eq:redChiSquared}
\end{equation}
where $N$ is the number of data points, $y_i$ are the experimental values of $P_n$ with statistical error $\delta_i$, calculated over $10^5$ measurements, and $s_i$ are the points simulated under the same conditions as the experimental data, using Eq.~(5) of the main text.
For each different magnetic field, we take into account different kinds of DD sequences in order to calculate the residuals of all of them at once. For example, at ($635\pm1$)~G we obtain the mean-squared-residuals from four datasets of AXY-8 together with one dataset of an Uhrig (UDD) sequence, as shown in Fig.~\ref{fig:axy8all} and Fig.~5(b)--main text, respectively. 
A second example is shown in Fig.~\ref{fig:simDat_redChiSqua}. , where the experimental data together with the simulation are shown for $B=(394\pm1)$~G. 

Performing this kind of analysis for all the investigated magnetic field intensities, we obtain $\chi_{\nu}^2$ with respect to the NSD obtained by measuring $T_2^L$, shown as squares in the inset of Fig.~4(c) of the main text. 
It is important to notice that the values of the mean-squared-residuals increase strongly for low magnetic fields, this will be addressed in the following section.
A summary of the DD sequences used to calculate the values of $\chi_{\nu}^2$ is shown in Table~\ref{tab:SummaryResidualsData}.

\begin{figure}
\includegraphics[width=\columnwidth]{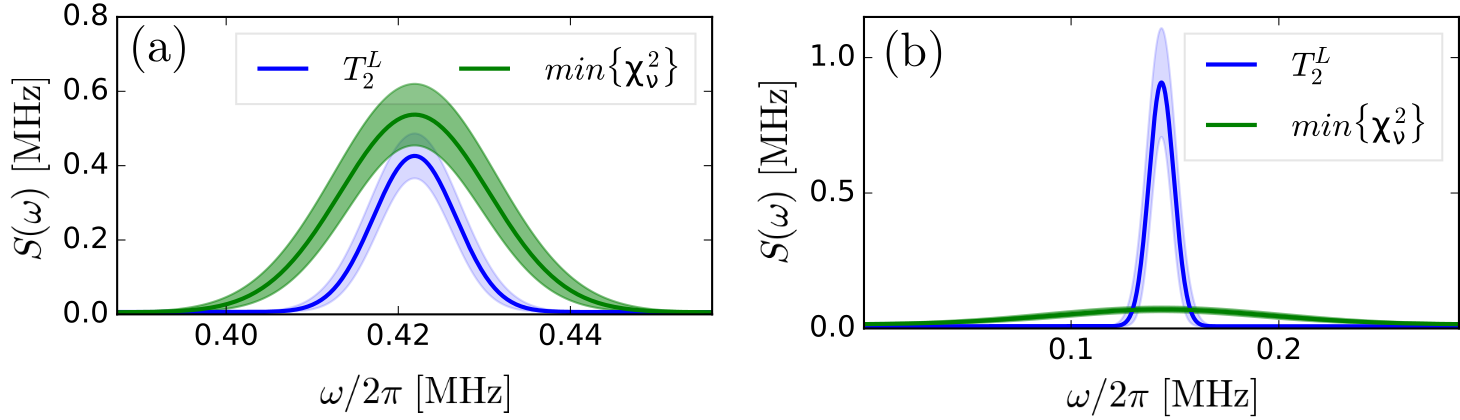}
\caption{NSD obtained from the measurement of $T_2^L$, as described in the main text (blue) and NSD obtained from the minimization of $\chi_\nu^2$. The shaded area represent the errorbars on the parameters that describe the Gaussian peak.
(a) For $B=(394\pm1)$~G -- weak coupling regime. The simulation using each versions of the NSD is shown on Fig.~\ref{fig:simDat_redChiSqua}.
(b) For $B=(132\pm1)$~G -- strong coupling regime. The simulation using each versions of the NSD is shown on Fig.5~(d,e) of the main text.
}
\label{fig:2NSDcomparison}
\end{figure}

\begin{table}\renewcommand{\arraystretch}{1.2}
\caption{Summary of the data sets used to extract the mean-squared-residuals, in terms of the external magnetic field. The $\chi_{\nu}^2$ values resulting from comparison between simulation and experiment are shown as squares in the inset of Fig.~4(c) of the main text. Note that these datasets doesn't include the experiments used to characterize the NSD using the $T_2^L$ method.
Notation: XY-8 is the equidistant sequence described in the main text, AXY-8 is the adaptive XY-8, UDD means Uhrig sequence, and SE stands for spin-echo.}
\begin{tabular}{clc}
\hline 
B-field [G]$\quad$ & Base DD sequence$\quad$ & number of pulses \\ 
\hline
\multirow{5}{*}{$635(1)$} & UDD & 32 \\
& AXY-8 ($r_m=1$) & 40 \\
& AXY-8 ($r_m=0.75$) & 40 \\
& AXY-8 ($r_m=0.5$) & 40 \\
& AXY-8 ($r_m=0.25$) & 40 \\
\hline
\multirow{5}{*}{$528(1)$} & XY-8 & 8 \\
& XY-8 & 24 \\
& XY-8 & 32 \\
& UDD & 32 \\
& XY-8 & 48 \\
\hline
\multirow{4}{*}{$394(1)$} & SE & 1 \\
& XY-8 & 8 \\
& YX-8 & 24 \\
& YX-8 & 32 \\
\hline
$309(1)$ & YX-8 & 24 \\
\hline
\multirow{2}{*}{$208(1)$} & AXY-4 ($r_m=0.75$) & 20\\ 
& YX-8 & 24 \\
\hline
\multirow{7}{*}{$132(1)$} & SE$^\dag$ & 1 \\
& AXY4$^\dag$ ($r_m=1$) & 20 \\
& AXY4$^\dag$ ($r_m=0.75$) & 20 \\
& UDD$^\dag$ & 32 \\
& UDD & 32 \\
& AXY8$^\dag$ ($r_m=1$) & 40 \\
& AXY8$^\dag$ ($r_m=0.75$) & 40 \\
\hline
\multirow{7}{*}{$78(1)$} & SE & 1 \\
& CPMG & 2 \\
& AXY4 ($r_m=1$) & 20 \\
& AXY4 ($r_m=0.75$) & 20 \\
& UDD & 32 \\
& AXY8 ($r_m=1$) & 40 \\
& AXY8 ($r_m=0.75$) & 40 \\
\hline
\multicolumn{3}{l}{$^\dag$ Using the $\{\ket{0},\ket{+1}\}$ qubit.}\\
\end{tabular} 
\label{tab:SummaryResidualsData}
\end{table}

\subsubsection{Strong coupling regime.}

As mentioned in the main text and in section~\ref{sec:classicalquantum}, in the strong coupling regime (low bias magnetic field) we expect the nuclear spin bath to be affected by the NV-spin back action dependently on the applied control sequence. 
Therefore, we find that we need more than one classical noise spectrum in order to achieve predictive results. 
We used two different methods to extract the NSD. 

\emph{Method 1:} We extract the NSD from the measurement of $T_{2L}$, as illustrated in the main text and in Sec.~\ref{Sec:NSDanalysis}. In the weak-coupling regime (high field), this method enables a spectrum reconstruction with a fully predictive capability. In the strong coupling regime, the reconstructed noise model reliably predicts the NV coherence only under DD sequence with low $n$, owing to decoherence.
For example, with $B\sim110$ G and $n\geq 24$ equidistant pulses, the collapse related to the $2^{nd}$ harmonic peak occurs at total times $T\geq \frac{T_1}{2}$. 
Any scheme to extract $T_2^L$ would require to use a small set of data with small $n$ values, thus entailing to relax the requirement of a large number of pulses in the definition of $T_2^L$. In addition, at low fields the collapses of the coherence function are overly deep, even for $n=1$, so that one can access only the tails of the NSD peak.
Still, the obtained noise model effectively predicts the NV spin dynamics under sequences with small $n$ (as Hahn-echo, shown in Fig.~5~(d) of main text) that have broad filter functions, less sensitive to the shape of the NSD. Under sequences associated to narrower filter functions, the NSD model is no longer suitable to make this prediction, as shown by Fig.~5~(e) of the main text and by the square-points in the inset of Fig.~4~(c) of the main text. 

\emph{Method 2:} For a given field, we simultaneously fit the $P_n$ datasets recorded under different control sequences, with simulation. 
For this fitting process we use as free parameters the NSD offset, amplitude and width, while we fix its center given by the Larmor frequency $\nu_L=B\gamma_{C_{13}}$, with $B$ the amplitude of the external magnetic field and $\gamma_{C_{13}}$ the nuclear spins gyromagnetic ratio. This method correctly describes the experimental data for sequences with $n\geq20$, but it fails for sequences with low number of pulses, like Hahn-echo or CPMG with $n=2$. An example of this is shown in Fig.5~(d,e) of the main text.

In the strong-coupling regime, these two methods result in NSD peak functions that are strongly different from each other, as shown in Fig.~\ref{fig:2NSDcomparison}~(b). 
Conversely, this is not the case for the weak-coupling regime, where the two noise models are slightly different (Fig.~\ref{fig:2NSDcomparison}~(a)), but both of them effectively predict the spin dynamics, as shown in Fig.~\ref{fig:simDat_redChiSqua}.

The triangles in the inset of Fig.~4~(c) of the main text, are calculated by evaluating the residuals using the first method for sequences with low $n$ values, and the second method for sequences with high number of pulses. This is referred to as the two-model picture in the main text.
We note that using the two-model picture results in values of $\chi_\nu^2$ comparable to the ones obtained from the $T_2^L$-based NSD measurement in the weak coupling regime.

\end{document}